\documentclass[aps,pre,twocolumn,groupedaddress]{revtex4}
\usepackage{amssymb}
\usepackage{amsbsy}
\usepackage{amsmath}
\usepackage{graphicx}

\begin{document}

\title{Intermittency and Velocity Fluctuations in Hopper Flows Prone to Clogging}

\author{C. C. Thomas and D. J. Durian}
\affiliation{Department of Physics and Astronomy, University of Pennsylvania, Philadelphia, PA 19104-6396, USA}

\date{\today}

\begin{abstract}
We experimentally study the dynamics of granular media in a discharging hopper. In such flows, there often appears to be a critical outlet size $D_c$ such that the flow never clogs for $D > D_c$. We report on the time-averaged velocity distributions, as well as temporal intermittency in the ensemble-averaged velocity of grains in a viewing window, for both $D < D_c$ and $D > D_c$, near and far from the outlet.  We characterize the velocity distributions by the standard deviation and the skewness of the distribution of vertical velocities. We propose a measure for intermittency based on the two-sample Kolmogorov-Smirnov $D_{KS}$-statistic for the velocity distributions as a function of time. We find that there is no discontinuity or kink in these various measures as a function of hole size. This result supports the proposition that there is no well-defined $D_c$ and that clogging is always possible. Furthermore, the intermittency time scale of the flow is set by the speed of the grains at the hopper exit. This latter finding is consistent with a model of clogging as the independent sampling for stable configurations at the exit with a rate set by the exiting grain speed [Thomas and Durian, Phys. Rev. Lett. (2015)].
\end{abstract}

\maketitle


\section{Introduction}

The gravity-driven flow of grains in an hourglass or hopper is an iconic granular phenomenon.  Fundamental issues of continued interest in granular physics today \cite{DuranBook, FranklinShattuck} include the shape of the coarse-grained velocity flow field \cite{Nedderman79, Tuzun79, Samadani99, Choi05, Garcimartin11}, the rate at which grains are discharged \cite{Beverloo, Nedderman82, MazaGM07, AguirrePRL2010, HiltonPRE11, JandaPRL12, RubioLargoPRL15, DunatungaJFM15}, and the susceptibility of the system to clogging \cite{HerrmannEPJE00, ToPRL01, Zuriguel05, Janda08, Mankoc09, Hannah_GM10, CloggingPD, ZuriguelPIP14, FofA}. The latter is usually quantified in terms of the average mass $\langle m\rangle$ or number of grains that is discharged before a clog occurs.  Experimentally, $\langle m\rangle$ is found to grow very rapidly with increasing hole size and may be fit to a power-law divergence in order to locate a clogging transition.  However, $\langle m\rangle$ may also be fit equally well to an exponential function, in which case there is no actual transition and clogging is in principle possible for any hole size.  If there truly is a transition, it ought to be possible to located it from above, i.e. by observing a critical change in some measured quantity as the hole size is decreased.  To our knowledge this has not been accomplished.  The closest is perhaps Ref.~\cite{Hannah_GM10}, where the nominal transition was bracketed by observing the stop and start angles at which a hopper with fixed hole size spontaneously clogs or unclogs as it is slowly tilted or untilted.  However such experiments depend on tilting rate.  Certainly, there is no discontinuity or kink in discharge rate versus hole size at the nominal transition \cite{Hannah_GM10, CloggingPD}.

The clogging transition, if it exists, is distinct from the jamming transition since the latter is for a spatially uniform system with no boundary effects.  Nevertheless, it seems likely that grains on the verge of clogging could display similarities to grains on the verge of jamming.  For example, there could be enhanced velocity fluctuations relative to the mean \cite{Menon97, LemieuxPRL00, Katsuragi10}, growing dynamical heterogeneities \cite{LucaBook}, or both \cite{Katsuragi10} as the hole size is decreased toward the transition.  A related but different question is to examine fluctuations versus time with a view toward predicting the imminence of clog formation for systems well within the clogging regime \cite{Tewari13}.  Here we explore behavior versus outlet size experimentally for a quasi-2D system of grains confined between clear parallel plates separated by a distance of about ten grain diameters, where discharge happens through a narrow slit at the bottom of the sample that extends across the full distance between the plates.  In particular, we use high-speed digital video particle-tracking techniques to measure fluctuations of the individual grain velocities, and intermittency of ensemble velocities, as a function of hole size both above and below the nominal clogging transition.  The results show that hopper flows susceptible to clogging have elevated fluctuations and are more strongly intermittent. However, this intermittency does not possess a time scale other than that set by average flow speed and grain size, i.e.\ it grows smoothly as the hole size is decreased through the transition, without any evidence of criticality.  This supports the notation that there is no actual well-defined clogging transition.

\section{Prior Experiments}

Well-known phenomenon involving fluctuations in granular hopper flow include silo quaking and ticking. These are typically observed with cohesive grains \cite{Mersch10} or where the interactions between the grains and the interstitial fluid \cite{TIcks, Bertho03} or the walls \cite{Borz11} are particularly strong.  Of more direct relevance for our work is study of intermittency in the mass discharge rate.  For example,  U\~{n}ac and co-workers found both fluctuations and characteristic time scales growing with decreasing hole size~\cite{Unac12}. However, Janda and co-authors found no such behavior \cite{JandaPRE09}.  Garcimart\'in and co-workers studied both the individual and ensemble velocity distributions as a function of hole size, but presented no systematic measurement of the fluctuation magnitude versus hole size~\cite{Garcimartin11}.  To our knowledge, the only peer-reviewed experimental work which explicitly reports on either the size of ensemble fluctuations or the time scale of the flow fluctuations throughout the bulk as a function of hole size is by Vivanco {\it et al.} for highly wedge-shaped hoppers~\cite{Vivanco12}. While this is an important geometry, it is more complicated, as these hoppers exhibit anomalous clogging statistics \cite{Saraf11}.  Thus there is need for the comprehensive characterization we report below.

\section{Experimental Methods}

For our work we use a quasi-2D hopper constructed with smooth, transparent, static-dissipative side walls.  The interior dimensions of the hopper are $3.8 \times 56$~cm$^2$, and it is typically filled to a height of at least 80~cm.  In this experimental regime there is no filling-height dependence of the flow \cite{Nedderman82}. The orifice is a rectangular slit at the bottom of the hopper with adjustable width $D$ and constant length $L = 3.8$~cm, running the full thickness of the hopper. The width of the slit, front and back, is measured with calipers both before and after experiments. In all cases the variation in the slit width is less than 0.1~mm.  We fix dowels on the hopper floor at the edges of the opening to inhibit the sliding of grains along the bottom. As a model granular medium suitable for imaging, we use $d=3.50 \pm 0.14$~mm monodisperse dry tapioca pearls. These grains are large enough that friction and hard-sphere repulsion are the only significant inter-granular interactions. Measuring the bulk density as $\rho_b = 0.69 \pm 0.01$~g/cm$^3$ and the density of an individual grain as $\rho_0 = 1.22 \pm 0.3$~g/cm$^3$, we estimate a packing fraction of $\phi = \rho_b/\rho_0 = 0.6 \pm 0.1$. This experimental setup is identical to one of those used in Ref.~\cite{FofA}.

\begin{figure}[!ht]
\includegraphics[width=3 in]{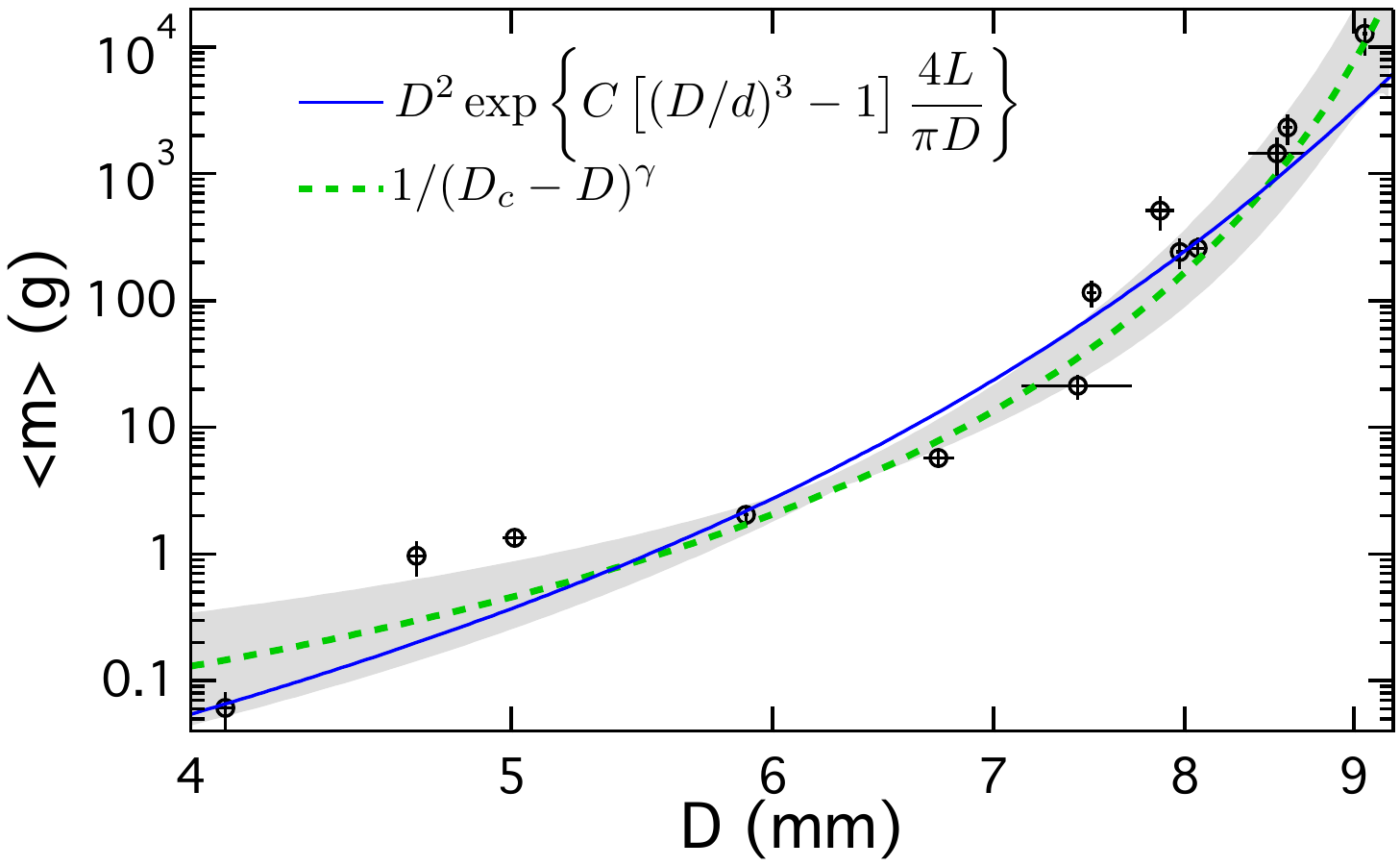}
\caption{(Color online). Average mass $\langle m \rangle$ discharged before a clog occurs, as a function of slit width $D$. Fitting to the clogging transition form of Eq.~(\ref{dc_eq}) we find $\gamma = 7.5 \pm 2.5$ and $D_c = 10.5 \pm 0.5\text{ mm}$, shown by the dashed green line. However, the approach of Ref.~\cite{FofA} suggests that $\langle m \rangle$ should follow the form of Eq.~(\ref{exp_eq}), shown by the solid blue line.}
\label{findDC}
\end{figure}

To locate the critical hole size $D_c$ for the nominal clogging transition, we follow standard procedure by fitting the average mass discharged before a clog occurs to a diverging power-law function:
\begin{equation}
	\langle m \rangle \propto \frac{1}{\left(D_c - D\right)^\gamma}.
\label{dc_eq}
\end{equation}
Data and fit are shown respectively as symbols and dashed line in Fig.~\ref{findDC}.  The fit given an exponent of $\gamma = 7.5 \pm 2.5$, consistent with prior observations.  And it gives the estimated location of the putative clogging transition as $D_c = 10.5 \pm 0.5$~mm.  This is an important number, that will be marked in several plots below.  The uncertainty in $\gamma$ and $D_c$, indicated by the grey band in Fig.~\ref{findDC} is a consequence of both the error in the variables and varying the fitting range. Equivalently, we may fit the data to a form suggested by Ref.~\cite{FofA}:
\begin{equation}
	\langle m \rangle = \frac{\pi}{4} \rho D^2 \ell \exp\left\{C\left[(D/d)^\alpha-1\right]\frac{4L}{\pi D}\right\},
\label{exp_eq}
\end{equation}
where $\rho = 0.69$~g/cm$^3$ is the bulk density of the tapioca. In Ref.~\cite{FofA}, we reported $\alpha = 3$, a sampling length $\ell = 0.75 d$, and $C =0.14 \pm 0.03$. Here we fix $\alpha = 3$ and find consistent values of $\ell/d = 0.9 \pm 0.2$ and $C = 0.117 \pm 0.004$. Fit to this latter form is overlaid as the solid line. The fits for both the exponential and the divergent form are very good: the ratio of $\chi^2$ for the exponential form to the divergent form is 1.05. We cannot therefore readily distinguish from such fits whether there exists a well-defined clogging transition for this hopper geometry.

\begin{figure}[!ht]
\includegraphics[width=2 in]{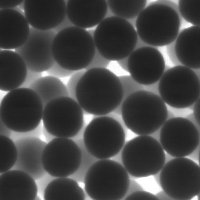}
\caption{Sample image of dry tapioca pearls in the quasi-2D hopper with back illumination. Only the grains nearest the camera, in sharp focus, are tracked and analyzed.  For scale, the average grain diameter is $d=3.5$~mm.}
\label{photo}
\end{figure}

\begin{figure}[!ht]
\includegraphics[width=3 in]{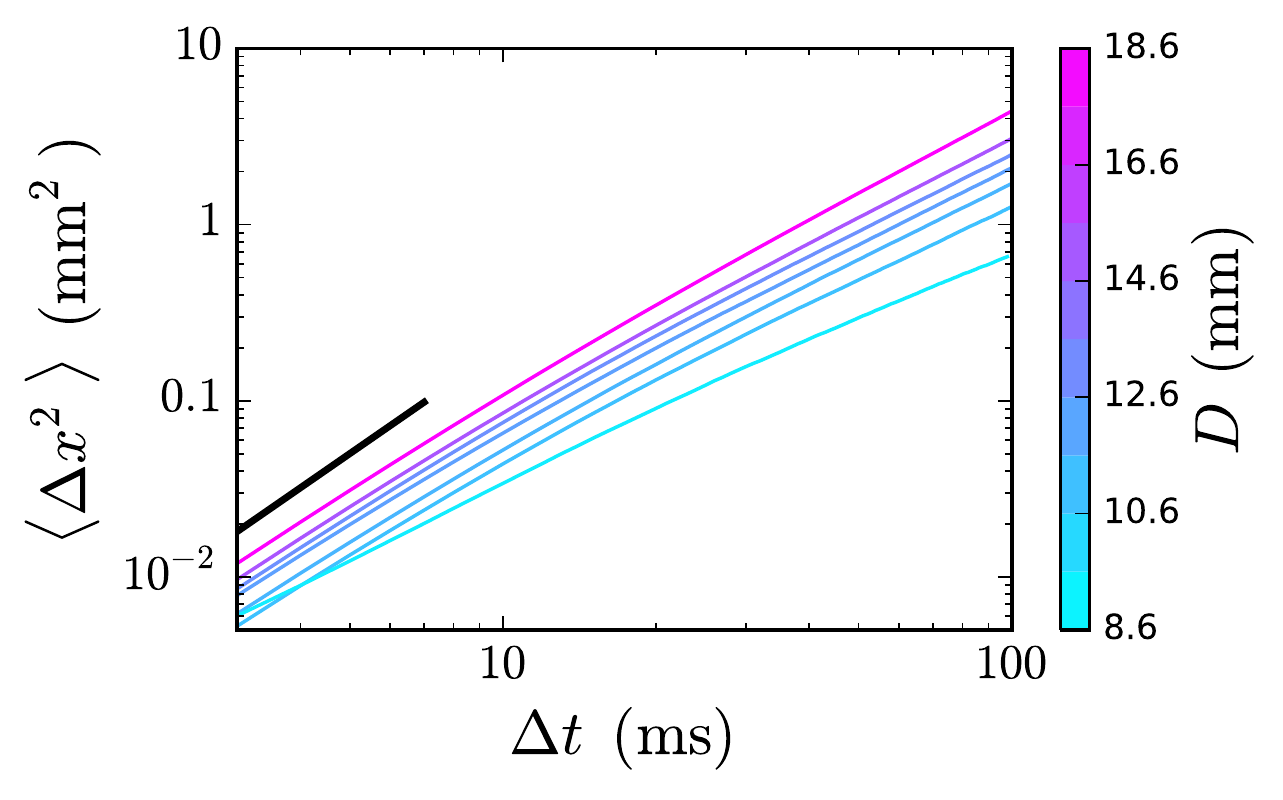}
\caption{(Color online) Mean squared displacement in the horizontal $x$-direction of grains near the exit. For all opening sizes, the motion is ballistic for short lag times $\Delta t$. The thick black line indicates $\langle \Delta x^2 \rangle \sim \Delta t^2$.}
\label{msd}
\end{figure}

We use a high-speed camera operating at 1~kHz frame rate to acquire images of the back-lit hopper. Only the grains at the wall nearest to the camera are within clear focus (see Fig.~\ref{photo}).  The positions of the particles are found to an accuracy of approximately 4 microns. The distance that the particles move between frames is far shorter than the typical distance between grains. We can therefore use the Crocker-Grier method to link the particle trajectories between frames \cite{Crocker96}.  As a demonstration, Fig.~\ref{msd} shows the mean-squared horizontal displacement $\langle \Delta x^2 \rangle$ versus delay time $\Delta t$ in a region near the exit.  Note that $\langle \Delta x^2 \rangle = v_x^2 \Delta t^2$ holds at short times.  Therefore, the spatial and temporal resolution is good enough to access the expected ballistic regime at short times and we thus have access to the true instantaneous particle speeds.

To find the instantaneous velocity $\boldsymbol{v}_i\left(t\right)$  of particle $i$ at time $t_0$, we fit $\boldsymbol{x}_i(t)$ in the range $-3$~ms~$< t-t_0 < 3$~ms to a second-order polynomial. Note that here we define $\boldsymbol{\hat{y}} \equiv -\boldsymbol{\hat{g}}$, and $(x,y) \equiv \boldsymbol{0}$ at the center of the slit. We restricted our data collection largely to a tall, narrow region centered above the slit, with $\left|x\right| < 2.5$~cm and $0 < y < 50$~cm.

\section{Velocity distributions}

\begin{figure}[!ht]
\includegraphics[width=3 in]{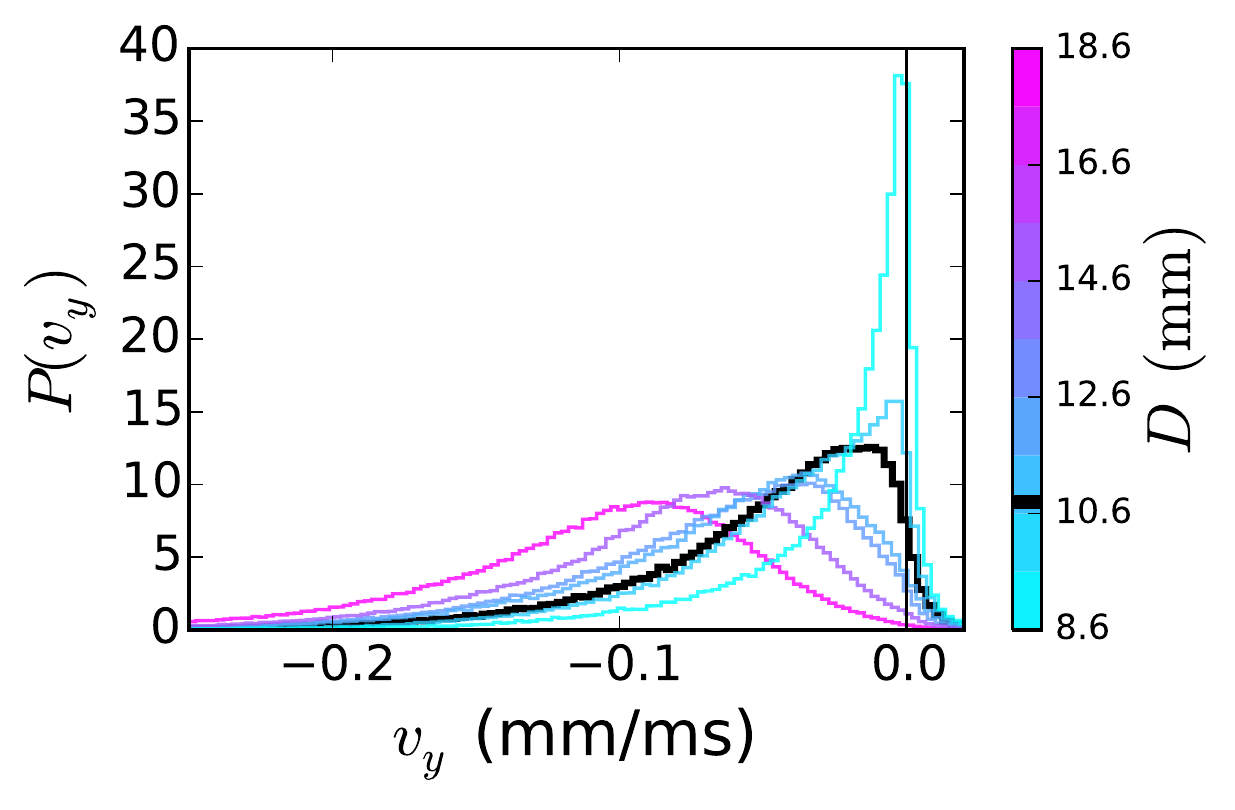}
\caption{(Color online) Velocity distribution of the grains in the vertical direction $y$ in a region near the exit. The distributions are broad and highly skewed in the downward direction.}
\label{vy_pdf}
\end{figure}

We begin by evaluating the distributions of these particle velocities, which depend sensitively both on the hopper opening size $D$ and the location of the grain within the hopper. Example velocity distributions in the vertical $y$-direction for the full range of opening sizes, in a region the the exit, are shown in Fig.~\ref{vy_pdf}. The black heavy curve indicates the distribution of $v_y$ when $D$ is near the putative clogging transition $D_c$. These are the velocity distributions for of the individual particles in a given region over all time. They are therefore distinct from the ensemble velocity statistics, discussed in detail in Section~\ref{sec:Intermittency}.

\subsection{Average flow}

\begin{figure*}[!ht]
\includegraphics[width=5 in]{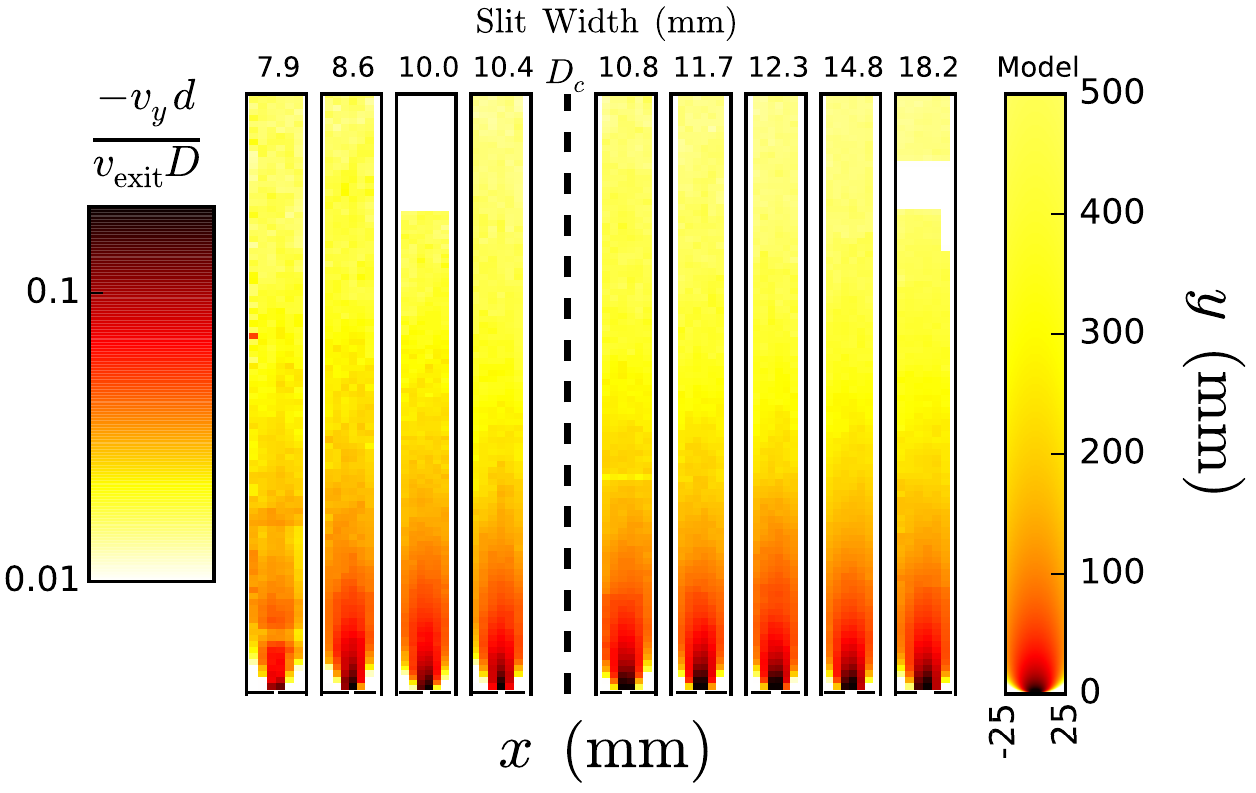}
\caption{(Color online) Map of the hydrodynamic average vertical velocity $v_{h,y}(x,y) \equiv \langle v_y(x,y) \rangle$ for hoppers of varying slit width $D$. Scaled by the speed of the exiting grains $v_{\rm{exit}}$ and $D$, the shape of the velocity field is identical for all hoppers. It is also in agreement with the prediction of the kinematic model, given by Eq.~(\ref{kinem_vy}) and shown on the far right with $b = 2d$. The estimate for $D_c = 10.5 \pm 0.5$~mm, from Fig.~\ref{findDC} and Eq.~(\ref{dc_eq}) is indicated. Binned regions shown are typically 7~mm$\times$ 5~mm.}
\label{vy_map}
\end{figure*}

We begin by considering the average of these velocity distributions.  We note how previous work has shown that the Beverloo equation for the average flow rate $W$ as a function of hole size $D$ works perfectly well for flows both above and below the clogging transition. In particular, Fig.~1 of Ref.~\cite{CloggingPD} demonstrates how there is no kink, or discontinuity in the first derivative, of $W(D)$ at the putative transition $D = D_c$. 

Not only is the discharge rate agnostic about the clogging transition, but the coarse-grained average (or hydrodynamic) granular velocities $\boldsymbol{v_h}(x,y)$ at various locations $(x,y)$ within the hopper also do not display any dependence on $D_c$.  Note that we determine the hydrodynamic velocity at $(x,y)$ by taking both the ensemble and the time average of \emph{all} grains within the bin at $(x,y)$ for all time: $\boldsymbol{v_h}(x,y) \equiv \langle \boldsymbol{v}_i(x,y;t) \rangle_{i,t}$.
The hydrodynamic velocity fields have long been understood to follow the empirical form \cite{Nedderman79, Tuzun79, Samadani99, Choi05, Garcimartin11}:
\begin{equation}
	\frac{v_{h,y}d}{v_{\text{exit}} D} = -\frac{d}{\sqrt{4 \pi b y}} \exp \left[-\left( \frac{x}{\sqrt{4 b y}}\right)^2 \right],
\label{kinem_vy}
\end{equation}
where $b$ is a length scale observed to typically range from $d$ to $3.5 d$. 

The hydrodynamic velocity field for our system is shown in Fig.~\ref{vy_map} for hoppers with slit widths $D$ both smaller and larger than $D_c = 10.5 \pm 0.5$~mm. The bins are rectangular regions typically 7~mm $\times$ 5 mm. As with the average flow rates, the shape is independent of $D$, with no difference of behavior above or below the clogging transition. For comparison with Eq.~(\ref{kinem_vy}), we plot the expectation when $b = 2d$ on the far right of Fig.~\ref{vy_map}.

\begin{figure}[!ht]
\includegraphics[width=3 in]{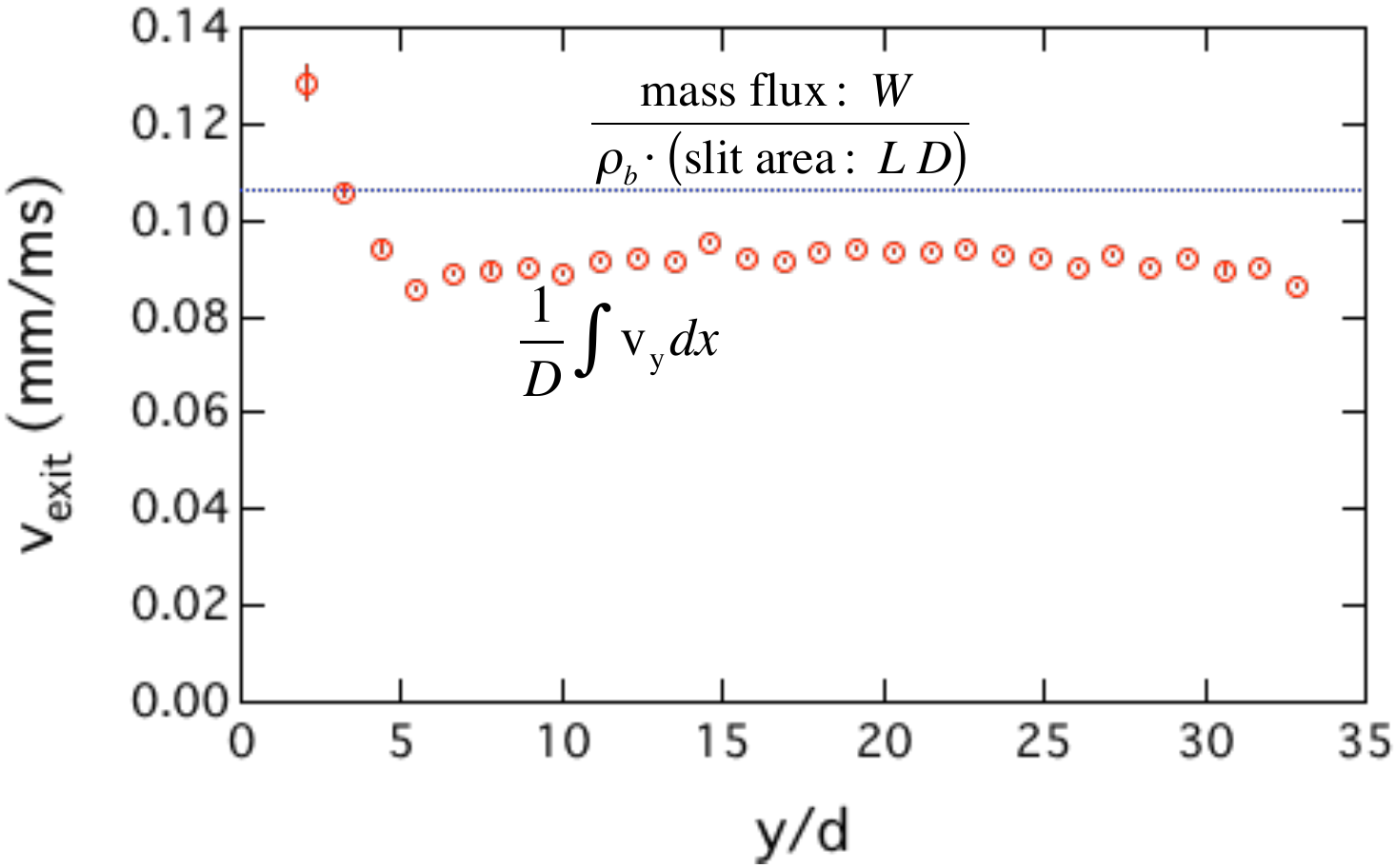}
\caption{(Color online). Exit velocity of the discharging grains, determined from either the bulk flow rate and Eq.~(\ref{vexit}) (dashed line) or the average velocity fields and Eq.~(\ref{vexit_field}) (red circles). The values of $v_{\text{exit}}$ for the imaged flows is within 15\% of that found from the bulk flow rate.}
\label{Quasi2dok}
\end{figure}

Although we are only imaging the layer of the grains at the wall, we confirm that the behavior here is fairly representative of the flow in the bulk. For this we determine the average speed of the exiting grains $v_{\text{exit}}$ for a slit width $D$ as:
\begin{equation}
	v_{\text{exit}} = W/(\rho L D)
\label{vexit}
\end{equation}
where $W$ is the average mass flow rate, $\rho$ is the bulk density of the tapioca, and $L$ is the length of the slit. At the orifice, this may be equivalently found by measuring the velocity directly: $v_{\text{exit}} = |\frac{1}{D} \int v_y(y=0) dx|$. By conservation of mass, we must have, for all $y$:
\begin{equation}
	v_{\text{exit}} = \left|\frac{1}{D} \int v_y(y) dx\right|.
\label{vexit_field}
\end{equation}
By comparing the values of $v_{\text{exit}}$ found from Eq.~(\ref{vexit}) and Eq.~(\ref{vexit_field}), we may evaluate how the measured flow fields compare with the bulk flow rate. This is shown in Fig.~\ref{Quasi2dok}. The flow rates imaged at the surface are within about $15\%$ of the flow rate within the bulk. We therefore conclude that friction with the walls does not significantly alter the flow patterns in this hopper.

\subsection{Velocity distribution moments}

\begin{figure*}[!ht]
\includegraphics[width=5 in]{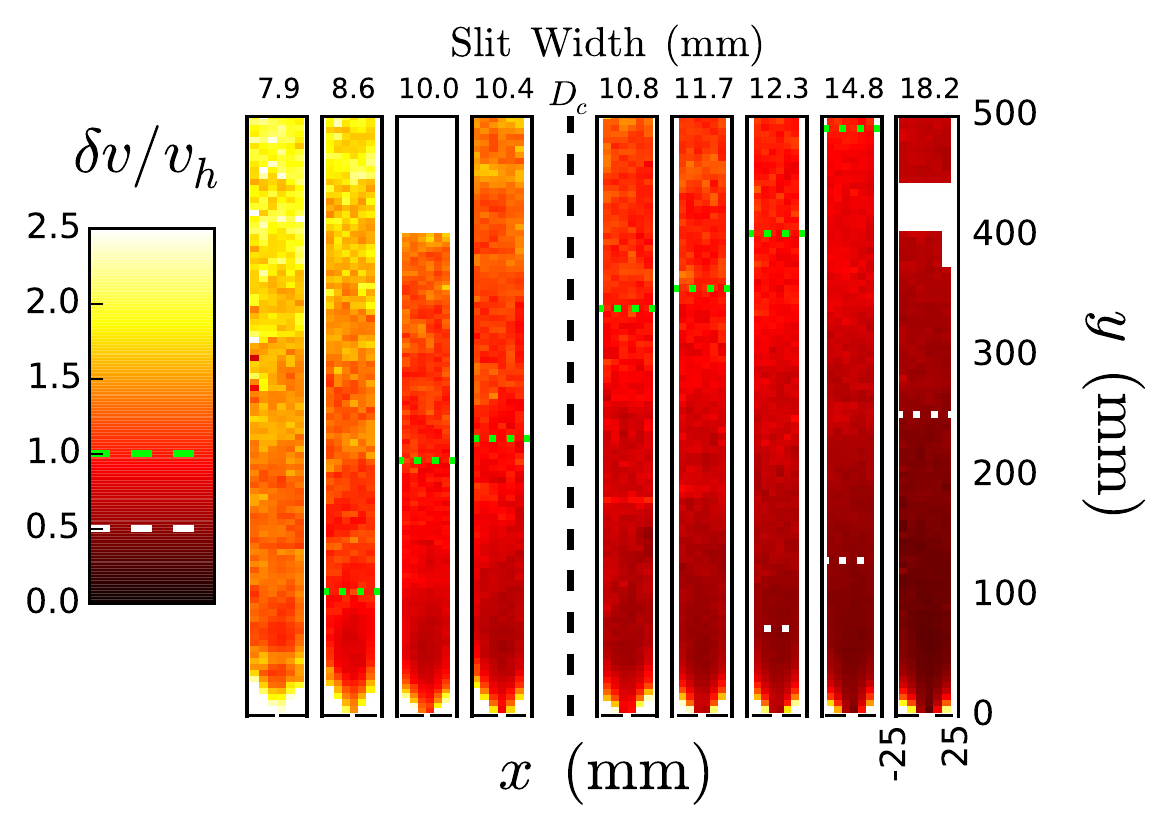}
\caption{(Color online) Map of the individual particle velocity fluctuations within a region over all time. Note that $\delta v^2 \equiv \sigma_{v_x}^2 + \sigma_{v_y}^2$ and $v_h^2 \equiv \langle v_x \rangle^2 + \langle v_y \rangle^2$. White lines indicate approximate locations of $\delta v = v_h/2$, while green lines indicate where $\delta v = v_h$. The relative velocity fluctuations increase in size for flows with smaller $D$, everywhere in the hopper. The estimate for $D_c = 10.5 \pm 0.5$~mm, from Fig.~\ref{findDC} and Eq.~(\ref{dc_eq}) is indicated.  Binned regions shown are typically 7~mm$\times$ 5~mm.}
\label{dv_map}
\end{figure*}

While the average flow behavior clearly does not provide evidence of the clogging transition, we may reasonably ask about the higher-order moments, including the standard deviation and the skewness of the velocity distribution.  Is there a significant change at $D = D_c$? Fig.~\ref{dv_map} demonstrates clearly that this is not so. Here, we measure $\delta v$ from the velocity distributions in both the horizontal and vertical directions: $\delta v^2 = \sigma_{v_x}^2 + \sigma_{v_y}^2$, where $\sigma$ is the standard deviation of \emph{all} the particle velocities within a particular binned region. (This is distinct from measuring the temporal fluctuations in the \emph{ensemble average} particle velocities.)  The associated granular temperature is $\sim m \delta v^2$.  Unlike the maps of the hydrodynamic velocity, there is a very significant hole-size dependence in $\delta v$. The fluctuations in the grain velocity are substantially larger for smaller slit widths. This occurs throughout the entire hopper, and is reminiscent of the $\delta v \sim v^{1/2}$ results obtained in Refs.~\cite{Menon97, LemieuxPRL00, Katsuragi10} using diffusing-wave spectroscopy. For the smallest slit width, $\delta v > v_h$ holds everywhere. However, this transition from low fluctuations to high fluctuations is smooth as a function of $D$. There is no signature of a clogging transition in the $\delta v$.

\begin{figure*}[!ht]
\includegraphics[width=5 in]{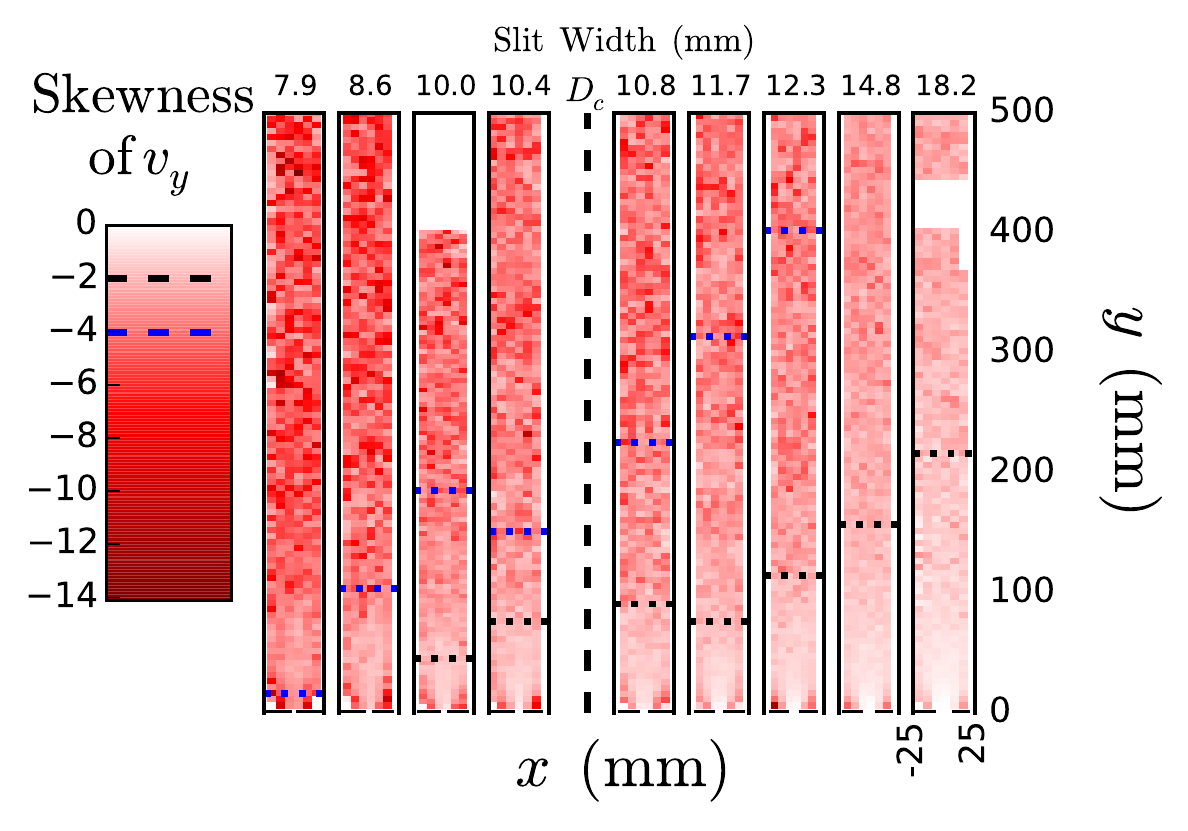}
\caption{(Color online) Skewness of $v_y$ at different locations within hoppers of varying slit width $D$. The fluctuations in the velocity are systematically more asymmetric for smaller $D$ and closer to the exit. The estimate for $D_c = 10.5 \pm 0.5$~mm, from Fig.~\ref{findDC} and Eq.~(\ref{dc_eq}) is indicated.  Black dashed lines indicate where the skewness $ = -2$, blue dashed lines where it is equal to $-4$. Binned regions shown are typically 7~mm$\times$5~mm.}
\label{skewvy_map}
\end{figure*}

Not only do hoppers more prone to clogging have larger fluctuations relative to the mean, but the fluctuations in the velocity are also more anisotropic. As evidence of this behavior, maps of the skewness in $v_y$ are shown in Fig.~\ref{skewvy_map}. Note that the sign of the skewness is always negative, that is, the distributions of the velocities are skewed in the downwards direction. As with the granular temperature, the magnitude of the skewness becomes larger everywhere in the hopper for smaller values of $D$. However, also like all the measures considered, there is no critical change in the skewness upon transitioning through $D_c$.

\section{Intermittency}
\label{sec:Intermittency}

\begin{figure*}[!ht]
\includegraphics[width=5 in]{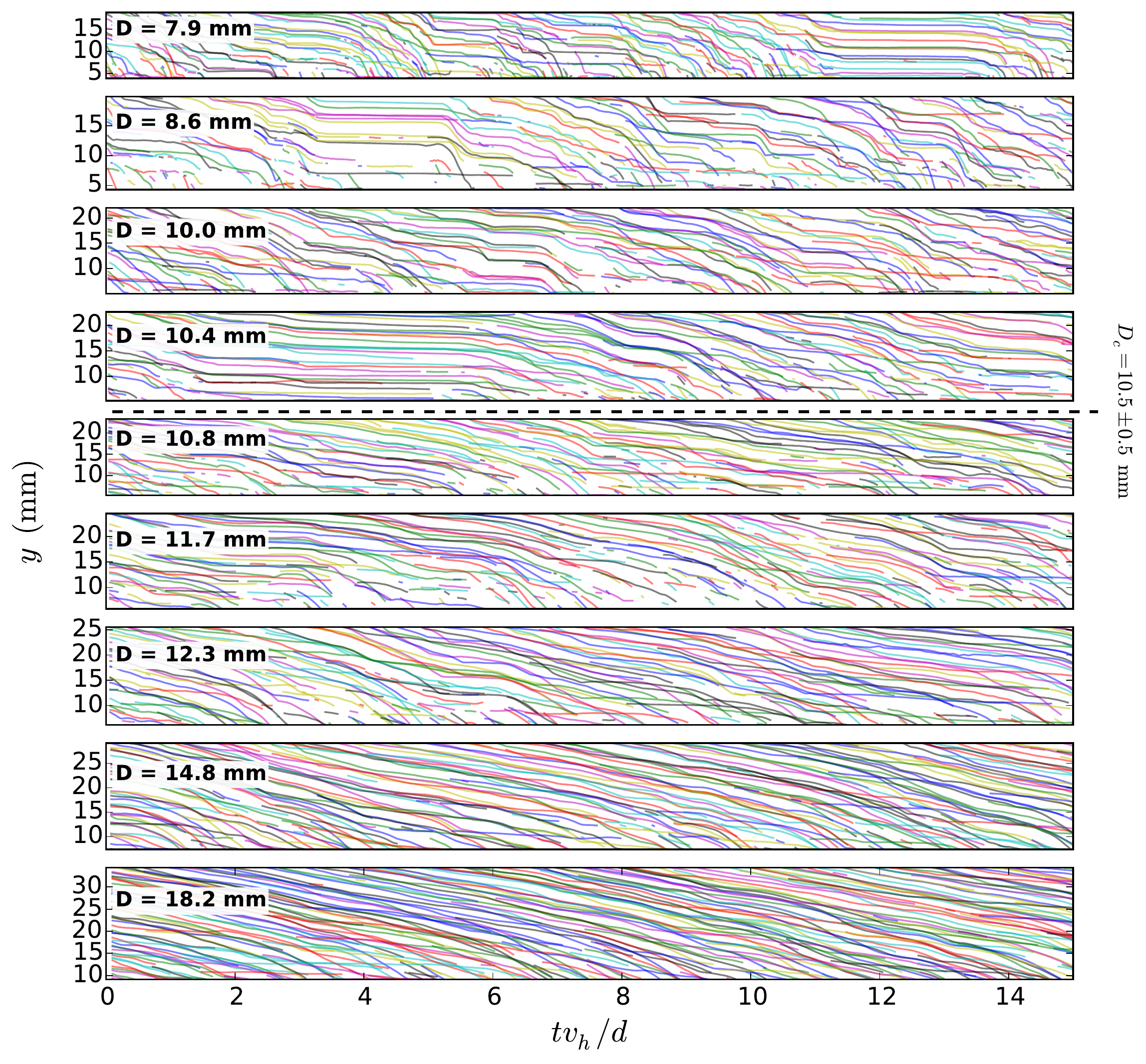}
\caption{(Color online) Vertical position of grains in the hopper near the exit over time $t$, where $t$ is scaled by the typical time $d/v_h$ for a grain to move a distance of its diameter $d$ downwards. All of the grains shown are within a square region of dimensions $\left(D + 2d\right) \times \left(D + 2d\right)$. The bottom of this region is a vertical distance $D/2$ from the slit and is centered about the center of the slit, $x = 0$. The dynamics of the grains vary substantially with slit width $D$: for small widths, the grains will often come to a rest for a period of time before flowing again. For large $D$, the ensemble speeds do not fluctuate as much and are never observed to approach zero.
}
\label{spacetime}
\end{figure*}

The previous measures of the flow considered only time-averaged statistics of individual particles. However, a very striking feature of slow granular hopper flow is the development of collectively intermittent dynamics as the hole size decreases. In particular, multiple grains in a viewing region speed up and slow down in tandem, more so for smaller holes.  This is illustrated in Fig.~\ref{spacetime}, where we plot vertical particle positions versus time.  We do so only for grains in a square viewing region directly near the exit: the bottom of the region is at $y = D/2$. The region size is set by $D$ as $\left(D + 2d\right) \times \left(D + 2d\right)$, and the region is centered about $x = 0$. Here, we scale time on the $x$-axis by the typical time $d/v_h$ that a grain translates by its own diameter. The difference in behavior between small $D$ and large $D$ is quite striking: for small $D$, many grains often come to a near-stop before the flow resumes again. Clogging would be an extreme instance in which the grains actually stop altogether, forever.  For large $D$, there is also significant collective behavior, but the ensemble velocity instead fluctuates between periods of slightly faster and slightly slower flow. Clearly the flow is more intermittent for small $D$ than for large $D$.  This can also be seen in movies of the flow near the exit for two different hopper sizes \cite{supp}.  Next, we consider several ways to characterize this intermittency.

\begin{figure*}[!ht]
\includegraphics[width=5 in]{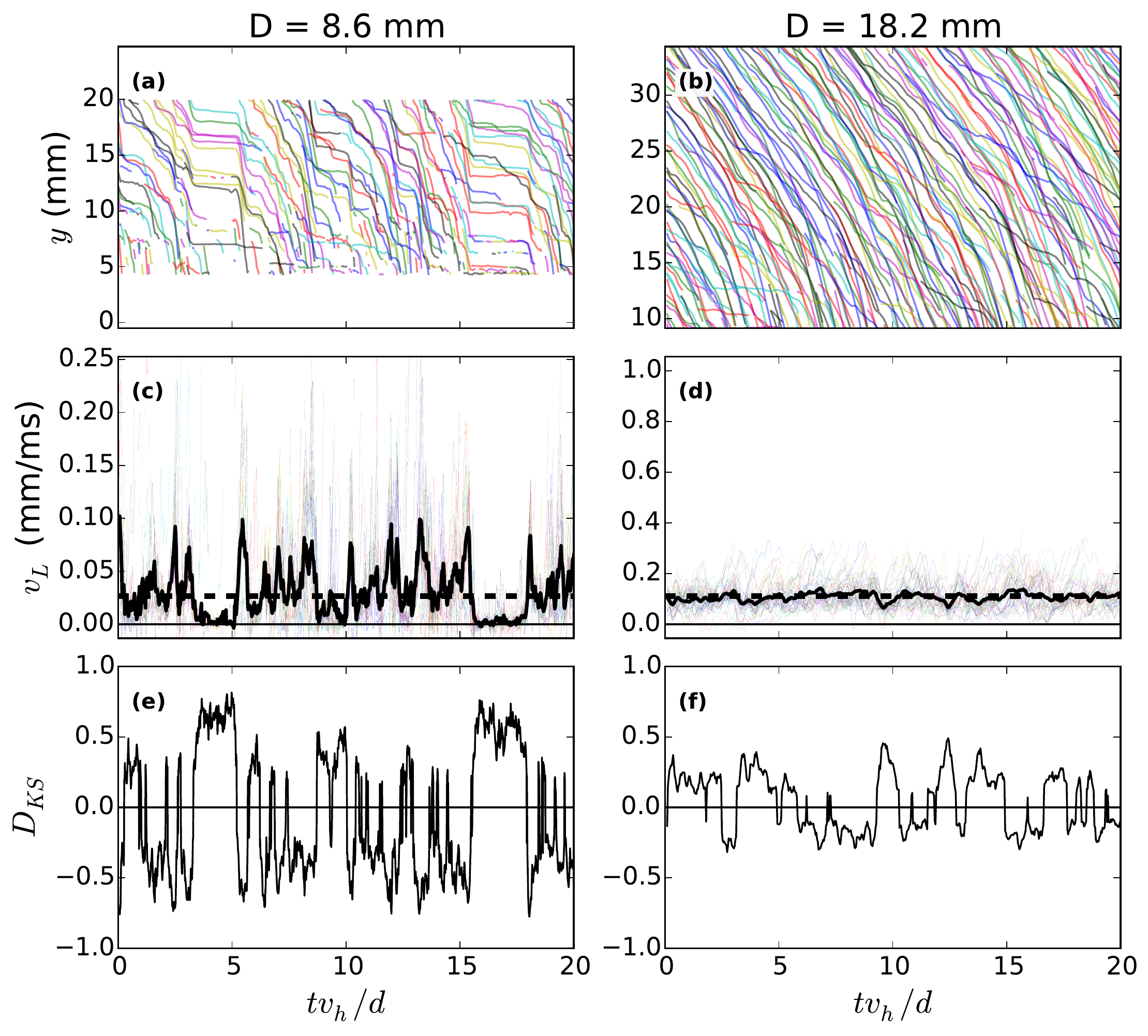}
\caption{(Color online) Behavior near the exit region for two hoppers with different slits widths of $D = 8.6$~mm (left) and $D = 18.2$~mm (right). Particles are in a square region near the exit with dimensions $\left(D + 2d\right) \times \left(D + 2d\right)$. 
(a,b) Vertical position $y$ of grains over time, as shown in Fig.~\ref{spacetime}. (c,d) The component $v_L$ of the individual grain velocities $\boldsymbol{v}_i\left(t\right)$ over time. The ensemble velocity $v_{E,L} \equiv \langle v_{i,L}\left(t\right) \rangle_i$ is overlaid as a thick black curve. The horizontal dashed line indicates the average $v_L$ over all time. (e,f) Signed Kolmogorov-Smirnov statistic $D_{KS}\left(t\right)$. The statistic characterizes the deviation of the velocity distributions at time $t$ with the velocity distribution for all time. When $D_{KS}\left(t\right) \approx 0$, the velocity distribution at time $t$ is very similar to the distribution over all time. When $D_{KS}\left(t\right) > 0$, the grains are moving slower than typical, and when $D_{KS}\left(t\right) < 0$, the grains are moving faster than typical. We therefore classify the behavior at time $t$ as ``fast" or ``slow" by the sign of $D_{KS}\left(t\right)$.
 }
\label{compare}
\end{figure*}

A simple measure to describe the collective behavior of the grains in the region is in terms of the ensemble velocity, $\boldsymbol{v_E}(t) \equiv \langle \boldsymbol{v_{i}}(t) \rangle_i$, where the average is taken over all $i$ particles in a region of interest at time $t$. Here, we consider the component of this ensemble velocity in the direction of the average flow, or the longitudinal velocity: $v_L = \boldsymbol{v} \cdot \boldsymbol{\hat{v}_h}$. In Fig.~\ref{compare}c,d, we display both the individual $v_{i,L}$ and the ensemble $v_{E,L}$ over time for a highly intermittent case (c) and a case with little intermittency (d). Not only is $\delta v/v_h$ larger, but the relative ensemble velocity fluctuations are larger for smaller $D$.

The ensemble-averaged velocity is helpful for describing the intermittency of the collective behavior, but it does not provide the full story. We wish to understand how the collective behavior as a whole differs from one time to another. For example, heap granular flow exhibits extreme intermittency with characteristic on/off time scales \cite{Lemieux00}.  We hypothesize that the velocity distributions during the ``pause events" seen in Fig.~\ref{compare}a,c are characterized by different velocity distributions than during regular flow.

To test this hypothesis, we calculate the Kolmogorov-Smirnov (K-S) statistic comparing the velocity distribution $P_t(v_{i,L})$ at time $t$ with the distribution over all time $P(v_{i,L})$ \cite{NumRec}. The two-sample Kolmogorov-Smirnov (KS) statistic is defined as the maximum distance between the sample cumulative distributions. When its magnitude is near unity, the distributions are very dissimilar. If it is near zero, then the distributions are similar.
We add an additional tweak to this measure by determining the \emph{signed} KS statistic $D_{KS}$. This is identical to the usual KS statistic described above, except that it is negative when the $v_{i,L}(t)$ distribution is greater than the distribution for all the data. The result is shown in Fig.~\ref{compare}e,f, respectively. Contrasting these two, it is clear that more intermittent flow is characterized by a greater heterogeneity of the velocity distributions over time.

\begin{figure}[!ht]
\includegraphics[width=3 in]{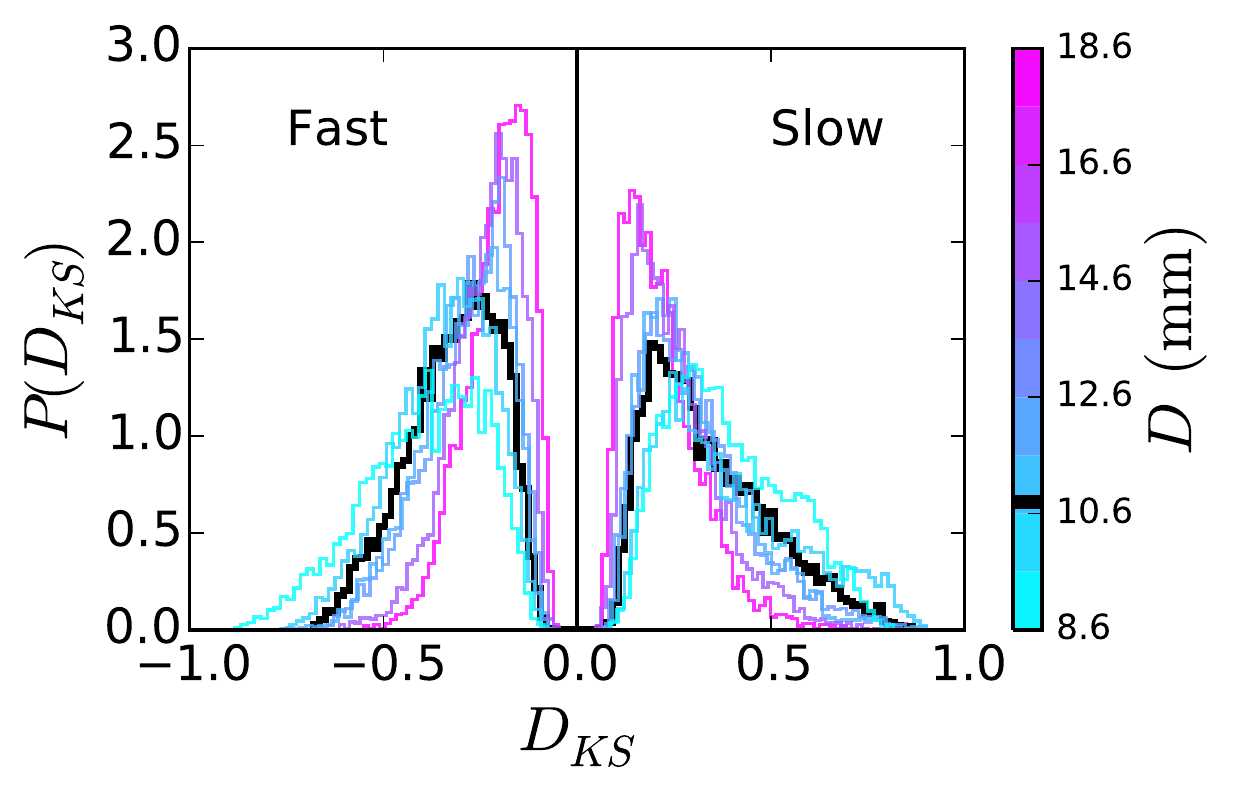}
\caption{(Color online) Probability distribution function of the signed Kolmogorov-Smirnov statistic $D_{KS}(t)$, for various slit widths $D$. Data are for the square region near the exit, as shown in Figs.~\ref{spacetime} and \ref{compare}, with dimensions $\left(D + 2d\right) \times \left(D + 2d\right)$. Times where $D_{KS}(t) < 0$ or $ >0$ can be classified as ``fast" or ``slow", respectively. The curve with $D \gtrapprox D_c$ is indicated as a black lines with heavier weight. The growing magnitude of $D_{KS}(t)$ for small $D$ indicates that the velocity distributions at any given time are less similar to the global velocity distribution, a signal of growing intermittency.}
\label{Dpdf}
\end{figure}

The intermittency as a function of opening size $D$ is further illuminated in Fig.~\ref{Dpdf}, which shows the the probability distribution of $D_{KS}$ for various slit widths $D$. When $D$ is smaller, the velocity distributions at any given time are more likely to deviate from the long-time velocity distribution, as seen by the larger distribution in the magnitudes of $D_{KS}$. Note also that for all slits there is a minimum value of $\left|D_{KS}\right|$: at all times $P_t(v_L)$ deviates significantly from $P(v_L)$. We can therefore consider the collective particle behavior at any time $t$ to be either ``fast" or ``slow", depending on the sign of $D_{KS}$.

\begin{figure}[!ht]
\includegraphics[width=3 in]{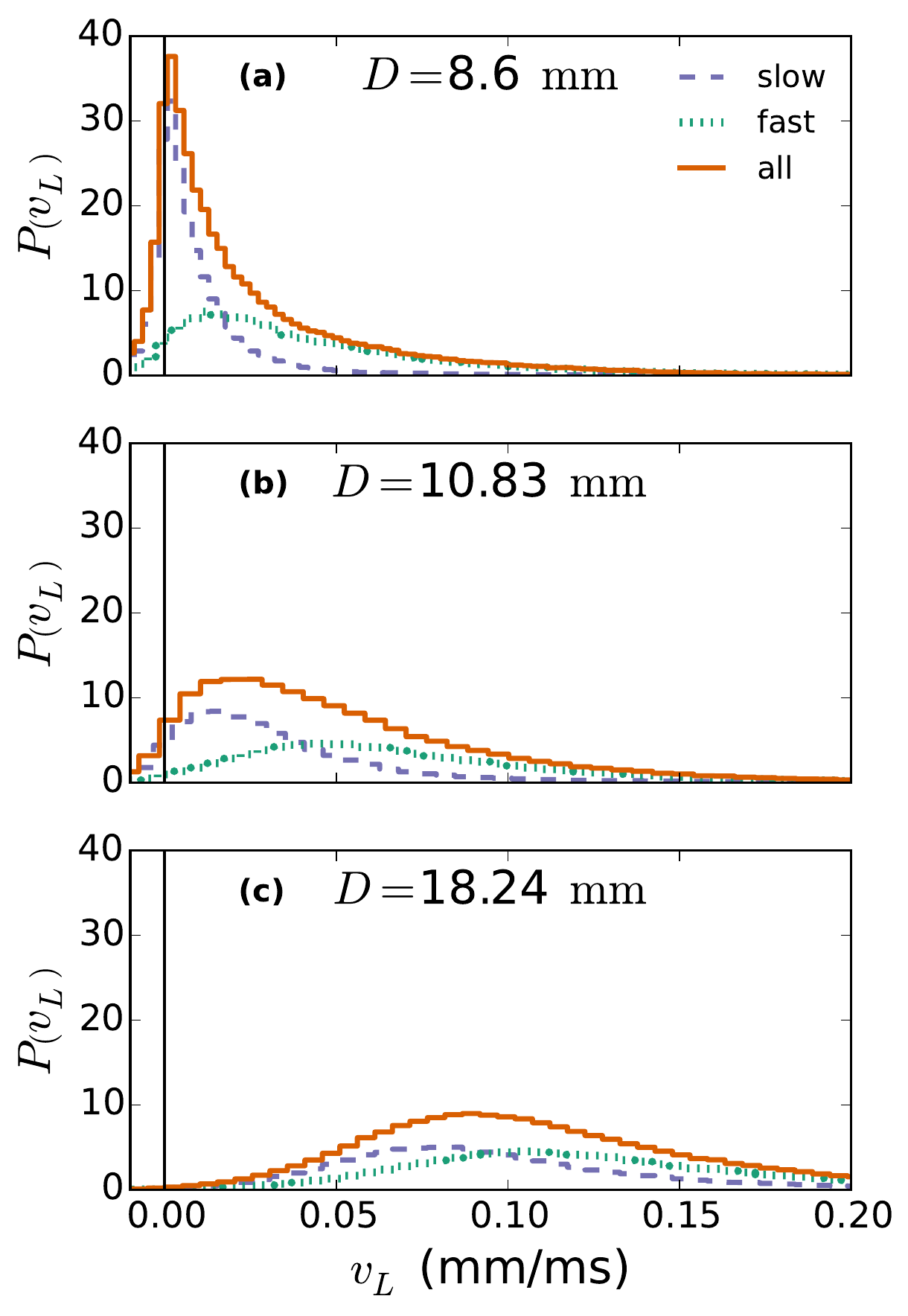}
\caption{(Color online) Distributions of the individual particle velocities $v_{i,L}(t)$ for several different slit widths $D$ in the region near the exit. The region has dimensions $\left(D + 2d\right) \times \left(D + 2d\right)$. Red solid curves indicate the distributions over \emph{all} time: they are characteristically more skewed for slower flows more prone to clogging. The distribution of the individual velocities during ``slow" times, when $D_{KS}(t) > 0$, are shown as dashed blue curves, and for ``fast" times, when $D_{KS}(t) < 0$ as dotted green curves. For large opening sizes $D$ the velocity distributions are similarly gaussian in character. However, for small $D$ the distributions during slow periods increasingly deviate from the fast velocity distributions, including larger probabilities of velocities that are zero or even opposite the direction of the mean flow ($v_L < 0$).}
\label{fast_slow}
\end{figure}

For any slit width, we classify a given particle velocity $v_{i,L}(t)$ as ``fast" or ``slow" if $D_{KS}(t) < 0 $ or $> 0$, respectively. The separate velocity distributions for the particles at fast or slow times are shown in Fig.~\ref{fast_slow} for several different $D$. When $D$ is small, $P(v_L)$ is very different for the fast and the slow cases. As seen in Fig.~\ref{compare}c, the ensemble velocity switches between ``flowing" and ``paused" states. However, for larger $D$, the distributions are much more similar, and flow typically switches instead between ``fast" and ``slow" states. 

\begin{figure}[!ht]
\includegraphics[width=3 in]{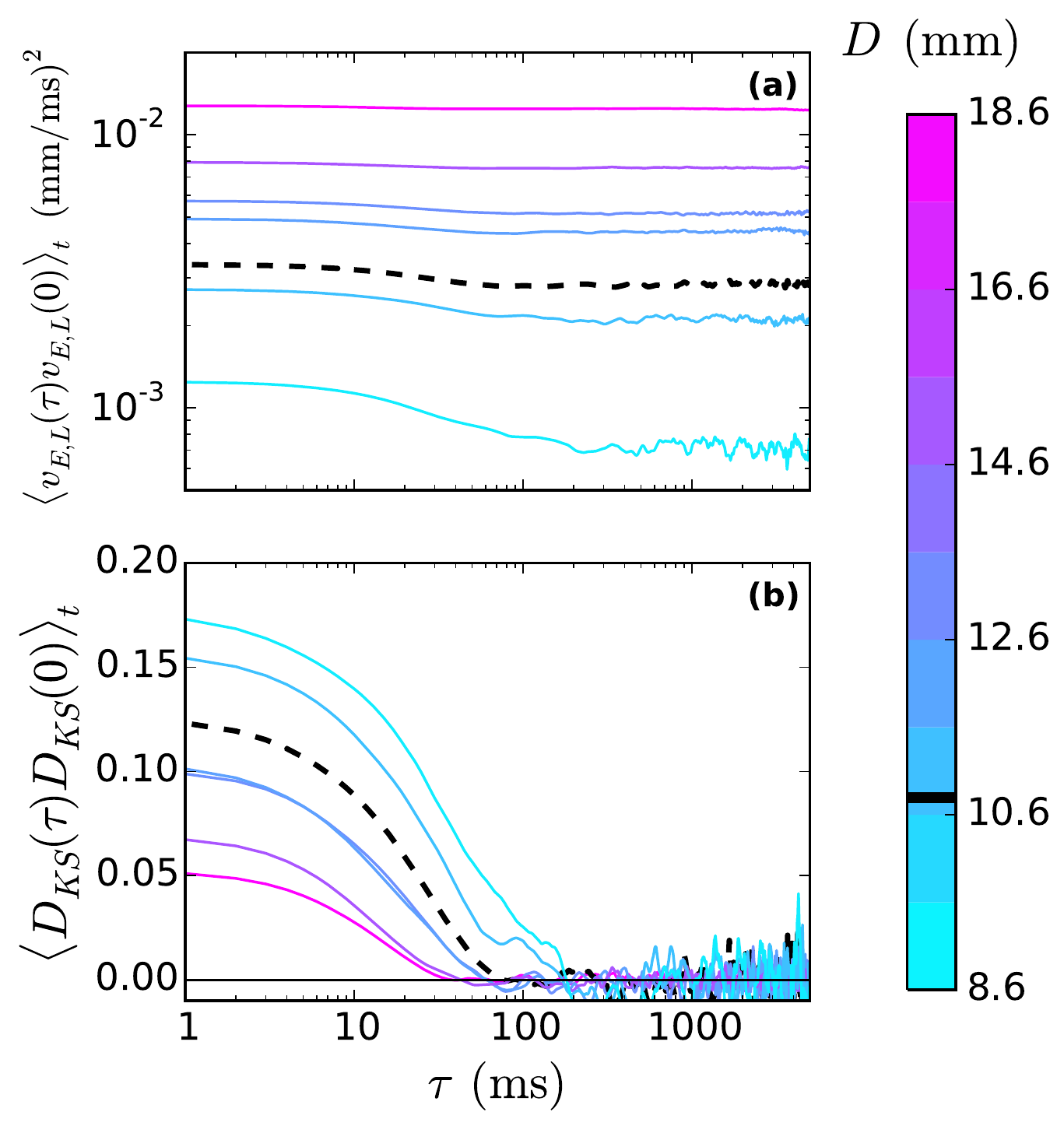}
\caption{(Color online) Autocorrelation functions of (a) ensemble velocity $v_{E,L}(t)$ in the direction of the mean flow and (b) the signed Kolmogorov-Smirnov statistic $D_{KS}(t)$, where examples of $v_{E,L}(t)$ and $D_{KS}(t)$ are shown in Fig.~\ref{compare}c,d and Fig.~\ref{compare}e,f, respectively. These are shown here only for the region near the exit, with dimensions $\left(D + 2d\right) \times \left(D + 2d\right)$. The autocorrelation function time scales grow slowly with decreasing $D$. The curve with $D \gtrapprox D_c$ is indicated by a black dashed line with heavier weight. The time scales and magnitude of the fluctuations are shown in more detail in Fig.~\ref{measures}.}
\label{acorr}
\end{figure}

\begin{figure}[!ht]
\includegraphics[width=3 in]{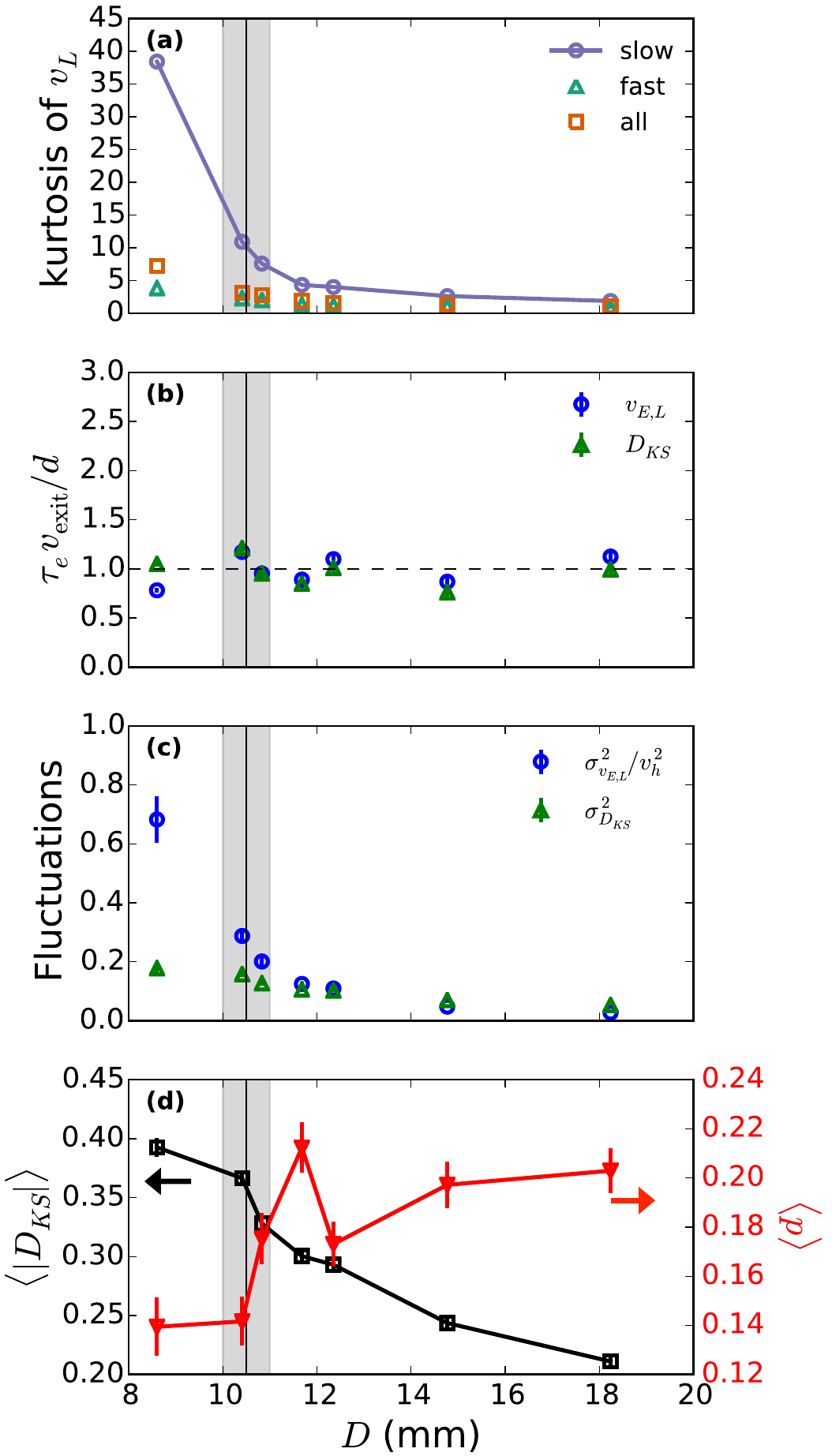}
\caption{(Color online) Signatures of intermittency for the square region near the exit of dimensions $\left(D + 2d\right) \times \left(D + 2d\right)$. The value of $D_c$, as found in Fig.~\ref{findDC} and Eq.~(\ref{dc_eq}) is displayed as a vertical line with a grey band indicating the confidence intervals of $D_c$. (a) Excess kurtosis of the longitudinal velocity during slow (blue circles), fast (green triangles), and all (red squares) times, as a function of slit width $D$. Fast or slow times are classified according to the sign of the signed Kolmogorov-Smirnov statistic $D_{KS}$. For smaller $D$, the difference between the shape of the slow or fast velocity distributions is much larger, and there is a more sharply pronounced difference between the fast and the slow states. Error bars in $y$ are smaller than the displayed data points. (b) Time for the autocorrelation functions (Fig.~\ref{acorr}) of the ensemble velocity $v_{E,L}(t)$ and the signed Kolmogorov-Smirnov statistic $D_{KS}(t)$ to decay $1/e$ of their initial values above the baseline, scaled by the time that exiting grains move by their own diameter. (c) Fluctuations of $v_{E,L}(t)$ and $D_{KS}(t)$, as determined by the variance of these quantities (alternatively, the difference between the $y$-intercept and baseline in Fig.~\ref{acorr}). (d) Average value of $D_{KS}$ (black squares, left axis) and significance (red triangles, right axis). Larger values of $\langle |D_{KS}| \rangle$ indicate more intermittent flows. The significance of $D_{KS}$ is found as the $p$-value, with average significance plotted on the right. Even though the sample exit regions are smaller for smaller slit widths, the significance of $|D_{KS}|$ is greater (smaller $p$).
}
\label{measures}
\end{figure}

Next, we identify the time scales associated with this intermittency. The temporal autocorrelation functions of $v_{E,L}(t)$ and $D_{KS}(t)$ are calculated and displayed in Fig.~\ref{acorr}(a) and Fig.~\ref{acorr}(b), respectively.   We define the time scale $\tau_e$ of the autocorrelation function as the value of $\tau$ at which the autocorrelation function has fallen $1/e$ of the distance from the value at $\tau = 0$ to the baseline. Dividing $\tau_e$ by the characteristic time of grain motion at the exit $d/v_{\text{exit}}$, we see in Fig.~\ref{measures}b that the intermittency time scale is unsurprisingly identical to the time scale of the exiting grain motion. There is no diverging time scale associated with intermittency. In fact, the time scale for intermittency is simply set by the sampling rate of the clogging process.

However, the magnitude of the intermittency, evident in the difference between the $y$-intercept and the baseline of the autocorrelation plots of Fig.~\ref{acorr}, does increase for flows more prone to clogging. This is illustrated in Fig.~\ref{measures}(c). However, note that there is no kink in this quantity in the clogging transition. 
Alternatively, intermittency could be quantified by the absolute value of $D_{KS}$ rather than its standard deviation. This measure has the advantage in that its significance can be readily evaluated, as detailed in Ref.~\cite{NumRec}.  The calculated $p-$ value is the probability of randomly measuring a value of $|D_{KS}|$ at least as large as the one observed. The average values of $|D_{KS}|$, as well as the average values of the $p$-values, where the averages for both are taken over all time, is shown in Fig.~\ref{measures}(d). As with $\sigma_{D_{KS}}^2$, $|D_{KS}|$ grows steadily with decreasing opening size. Additionally, the $p$-values for $|D_{KS}|$ are greater for smaller $D$ as well, indicating that the growth of $|D_{KS}|$ is not a systematic effect due to smaller number of grains in the sampled region.

Finally, we may also describe intermittency by examining how the velocity distribution of grains during the ``slow" events changes with $D$ (blue dashed lines in Fig.~\ref{fast_slow}). We do this by calculating the excess kurtosis of these distributions, and plot in Fig.~\ref{measures}(a). The slow velocity distributions deviate substantially from Gaussian as $D$ decreases. However, as with the fluctuations, there is no signature of $D_c$ in this behavior.

\section{Intermittency and height dependence}

\begin{figure}[!ht]
\includegraphics[width=3 in]{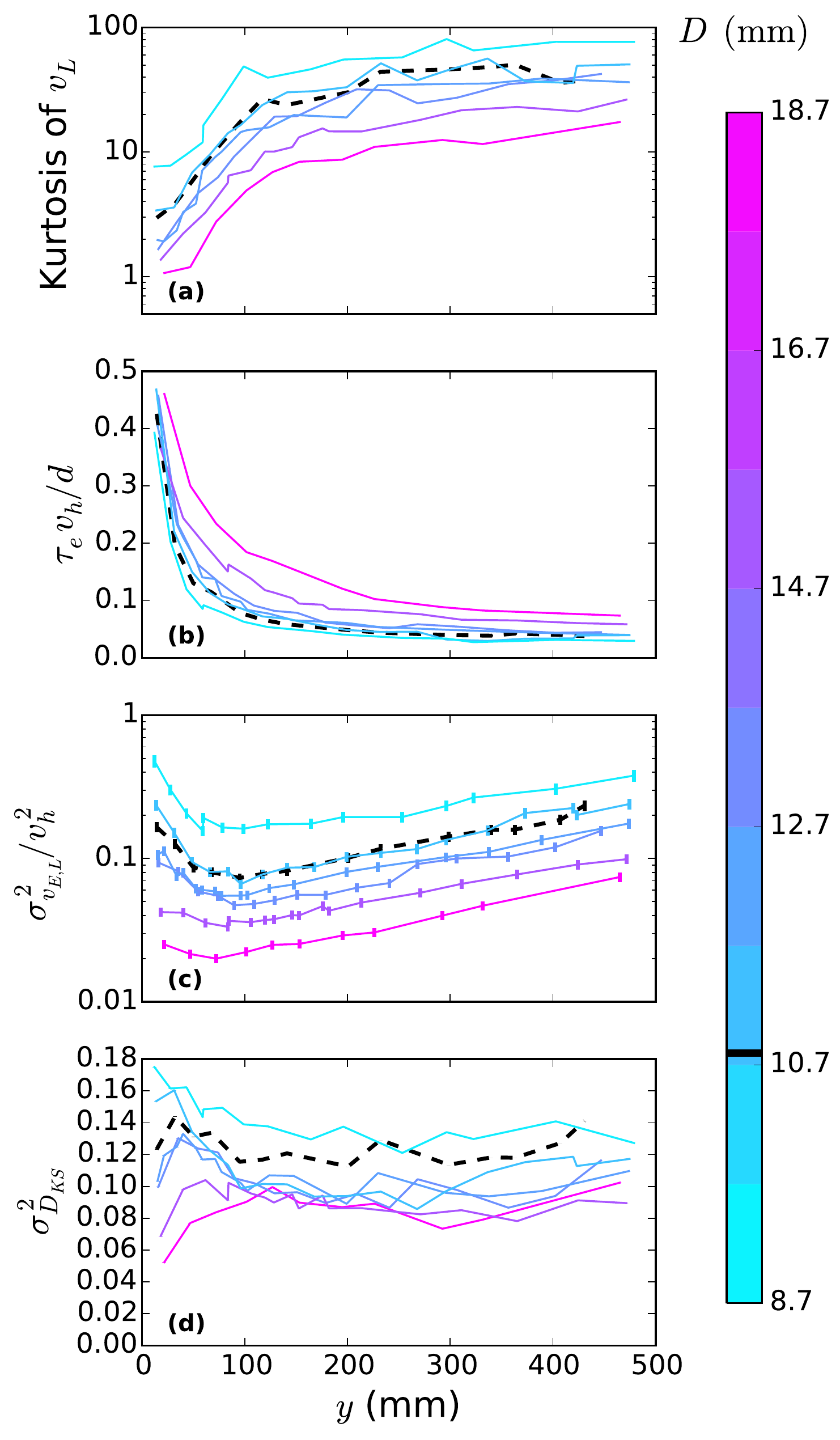}
\caption{(Color online) Plots of various quantities for regions within the hoppers as a function of vertical position $y$ and hopper opening width $D$. The curves with $D \gtrapprox D_c$ are indicated by black dashed lines with heavier weights.
(a) Excess kurtosis of the longitudinal velocity $v_L$ distributions, as a function of opening size $D$ and position in the hopper $y$. 
(b) Autocorrelation time scale $\tau_e$ for $D_{KS}(t)$, scaled by the average time it takes for a grain to shift by its own diameter $d$ at location $y$. Relative to the time scales of the average flow speed $v_h$, the intermittency time scale grows for small $y$ and large $D$. 
(c) Fluctuations in the ensemble velocity $v_{E,L}$ scaled by the hydrodynamic velocity $v_h$, for all $y$. 
(d) Fluctuations in $D_{KS}$. The fluctuations in the signed Kolmogorov-Smirnov statistic follow a different pattern for small $D$ than for large $D$: for hopper highly prone to clogging, the most intermittent flow is near the exit. For hoppers unlikely to clog, the most intermittent flow is far from the exit.
}
\label{all_y_measures}
\end{figure}

The flow near the hopper exit is more intermittent when it is more prone to clogging. However, the intermittency also varies throughout the hopper. Since width of the flowing region grows with height, the flows ought to be slower/more-intermittent higher up and faster/more-smooth near the exit.  Perhaps there is a boundary between intermittent and smooth regions that moves down toward the outlet as the size decreases toward a clogging transition?  To investigate, we evaluate large rectangular regions centered about $x = 0$. These are not the same bin sizes as displayed in the heat maps; in order to have sufficient statistics, we require that the average number of sampled configurations be approximately 400. We plot the kurtosis of the longitudinal velocity $v_L$ in Fig.~\ref{all_y_measures}(a). Following a similar trend as that shown in Fig.~\ref{measures}, we now see that the non-gaussian character of the velocity distributions grows everywhere in the hopper as $D \rightarrow d$. We also evaluate the autocorrelation time scale $\tau_e$ for $D_{KS}$ at different $y$, and scale the result by $d/v_h(y)$. Shown in Fig.~\ref{all_y_measures}(b), this demonstrates (in parallel with Fig.~\ref{measures}(b)) that nowhere is there an intermittency time scale longer than the inverse of the local average flow rate. As with the kurtosis of $v_L$, we find that the relative fluctuations $\sigma_{v_{E,L}}^2$ are a monotonic function of the slit width $D$ (Fig.~\ref{all_y_measures}(c)). 

The intermittency measure $\sigma_{D_{KS}}^2$ is displayed in Fig.~\ref{all_y_measures}(d). Here, we see clearly that the intermittency grows with the increasing likelihood of clogging.  However, the height dependence of $\sigma_{D_{KS}}^2$ is very different than that of $\sigma_{v_{E,L}}^2/v_h^2$.
For most hoppers, the flow is most intermittent near the exit, decreases with increasing $y$, and plateaus far from the aperture.
This demonstrates that the Kolmogorov-Smirnov statistic is a unique intermittency measurement which provides information not accessible from the instantaneous velocity distributions such as the kurtosis, skewness, or $\delta v/v_h$. Furthermore, the change in intermittency with height is not analogous to the change of intermittency with opening size. Both $\sigma_{v_{E,L}}^2/v_h^2$ and $\sigma_{D_{KS}}^2$ grow with decreasing $D$. However, they are not monotonic functions of $y$.

\section{Conclusion}

In conclusion, we have demonstrated that there is no critical change in behavior in granular flow near the putative clogging transition.  Relative velocity fluctuations $\delta v/v_h$, skewness in velocity, and intermittency magnitude $\sigma_{D_{KS}}$ all grow for flows more prone to clogging. However, there is no signature of the clogging transition in the variation any of these measures with hole size. This supports our suggestion in Ref.~\cite{FofA} that there is no well-defined clogging transition, and that $D_c$ is simply a hole size at which the probability for a flow to clog on laboratory time scales disappears.

We have proposed what we believe is a useful new measure for quantifying intermittency and the tendency for flows to fluctuate between different parent velocity distributions. While in this case we simply considered the alternation between fast and slow flows, this method can easily be generalized to encompass more complicated cases.

Finally, we find no evidence of a distinct intermittency time scale anywhere in the hopper, either in the ensemble average flow rate $v_{E,L}(t)$ or the intermittency $D_{KS}(t)$. The time scales for both of these quantities to is identical. For both, the longest time scale anywhere in a hopper of a given hole size $D$ is $d/v_{\rm{exit}}$, the rate at which grains move by their own diameter in the exit region. Significantly, there is no evidence of a diverging time scale, either as $D \rightarrow D_c$ or $D \rightarrow d$. Hoppers with smaller hole sizes are not closer to jamming. Rather, hoppers with smaller hole sizes are more likely to fall into a jammed state due to the smaller number of grains at the exit that are required to be ``pre-clogged".

We previously demonstrated that clogging is a Poisson sampling process independent over times greater $\tau_0 \sim d/v_{\rm{exit}}$. Together with the lack of a long time-scale for intermittency, this means that one cannot predict a clog with advance notice greater than $\tau_0$ by investigating the system dynamics. However, these results are for flows whose intermittency is set by the sampling behavior at the exit. It would be instructive to expand this analysis to cases with other sources of intermittency, for example, where interaction between the grains and the interstitial fluid contribute to intermittency. Do these other cases of intermittency coupling provide the necessary memory in the system to break the Poissonian nature of the sampling? Such explorations will further our general understanding of intermittent phenomena in systems near jamming in general and near clogging in particular.

\begin{acknowledgments}
This work was supported by the National Science Foundation through Grant No. DMR-1305199.
\end{acknowledgments}

\bibliography{FlowingRefs}

\end{document}